\documentclass[journal = ancac3]{achemso}
\setkeys{acs}{usetitle = true}

\usepackage{graphicx}
\usepackage{color}
\usepackage{amsfonts}

\title{Spatially Resolved Electronic Properties of Single-Layer WS$_2$ on Transition Metal Oxides}
\author{S{\o}ren~Ulstrup}
\affiliation{Advanced Light Source, E. O. Lawrence Berkeley National Laboratory, Berkeley,
California 94720, USA}
\altaffiliation{These authors contributed equally to the work}
\email{sulstrup@lbl.gov} 
\author{Jyoti Katoch}
\affiliation{Department of Physics, The Ohio State University, Columbus, OH 43210, USA}
\altaffiliation{These authors contributed equally to the work}
\author{Roland J. Koch}
\author{Daniel Schwarz}
\affiliation{Advanced Light Source, E. O. Lawrence Berkeley National Laboratory, Berkeley,
California 94720, USA}
\author{Simranjeet Singh}
\affiliation{Department of Physics, The Ohio State University, Columbus, OH 43210, USA}
\author{Kathleen M. McCreary}
\affiliation{Naval Research Laboratory, Washington DC 20375, USA}
\author{Hyang Keun Yoo}
\affiliation{Advanced Light Source, E. O. Lawrence Berkeley National Laboratory, Berkeley,
California 94720, USA}
\author{Jinsong Xu}
\affiliation{Department of Physics, The Ohio State University, Columbus, OH 43210, USA}
\author{Berend T. Jonker}
\affiliation{Naval Research Laboratory, Washington DC 20375, USA}
\author{Roland K. Kawakami}
\affiliation{Department of Physics, The Ohio State University, Columbus, OH 43210, USA}
\author{Aaron Bostwick}
\affiliation{Advanced Light Source, E. O. Lawrence Berkeley National Laboratory, Berkeley,
California 94720, USA}
\author{Eli Rotenberg}
\affiliation{Advanced Light Source, E. O. Lawrence Berkeley National Laboratory, Berkeley,
California 94720, USA}
\author{Chris Jozwiak}
\affiliation{Advanced Light Source, E. O. Lawrence Berkeley National Laboratory, Berkeley,
California 94720, USA}
\email{cmjozwiak@lbl.gov}

\begin{document}

\newpage

\begin{abstract}
There is a substantial interest in the heterostructures of semiconducting transition metal dichalcogenides (TMDCs) amongst each other or with arbitrary materials, through which the control of the chemical, structural, electronic, spintronic, and optical properties can lead to a change in device paradigms.  A critical need is to understand the interface between TMDCs and insulating substrates, for example high-$\kappa$ dielectrics, which can strongly impact the electronic properties such as the optical gap.  Here we show that the chemical and electronic properties of the single-layer (SL) TMDC, WS$_2$, can be transferred onto high-$\kappa$ transition metal oxide substrates TiO$_2$ and SrTiO$_3$.  The resulting samples are much more suitable for measuring their electronic and chemical structures with angle-resolved photoemission than their native-grown SiO$_2$ substrates.  We probe the WS$_2$ on the micron scale across 100-micron flakes, and find that the occupied electronic structure is exactly as predicted for freestanding SL WS$_2$ with a strong spin-orbit splitting of 420~meV and a direct band gap at the valence band maximum. Our results suggest that TMDCs can be combined with arbitrary multi-functional oxides, which may introduce alternative means of controlling the optoelectronic properties of such materials.\\
\\
KEYWORDS: Spatially-resolved photoemission, PEEM, ARPES, transition metal dichalcogenides, WS$_2$, high-$\kappa$ oxides.
\end{abstract}

\newpage

\maketitle

The isolation of single-layer (SL) semiconducting transition metal dichalcogenides (TMDCs)\cite{Novoselov2005, Ramakrishna2010} has enabled truly two-dimensional (2D) transistors and optoelectronic devices \cite{radisavljevic2011,wang2012,jariwala2013,jariwala2014,sdas2015,Bhimanapati2015}, as well as artificial heterostructures with interesting physical properties \cite{Geim:2013aa,lotsch2015,jariwala2016}. Among this family of TMDCs, which encompasses the 2H-polytypes of the four materials MoS$_2$, MoSe$_2$, WS$_2$ and WSe$_2$, the electronic properties of SL WS$_2$ are considered the most promising due to a combination of particularly strong spin-orbit coupling \cite{ramasl2012}, a small effective mass of the upper valence band maximum (VBM) \cite{Shi2013,Dendzik2015} and a relatively high mobility limited by electron-phonon coupling \cite{Ovchinnikov2014,Zhang2014}. Additionally, luminescence measurements of the pronounced A and B excitons indicate that WS$_2$ transforms into a direct gap semiconductor in the SL limit with the position of its VBM shifting to the $\bar{K}$-point in the corner of the Brillouin zone (BZ) of the material \cite{Zhao2013}, similarly as for MoS$_2$, MoSe$_2$ and WSe$_2$ \cite{makatomically2010,miwaelectronic2015,zhangdirect2014,Zhang2016}. Consequently, SL WS$_2$ supports strong light-matter interactions that lead to improved light absorption compared to standard photovoltaic devices \cite{britnellstrong2013} and exotic many-body effects such as dark excitons \cite{Ye2014}. 

The electronic properties of SL TMDCs are known to be extremely sensitive towards the properties of the supporting substrate or contact materials \cite{Ugeda2014,Ovchinnikov2014}. Depending on the dielectric screening by the substrate \cite{Ugeda2014,Antonija-Grubisic-Cabo:2015aa,Bruix:2016} and the number of free carriers induced either \textit{via} electrical or optical doping \cite{Chernikov:2015ab,Chernikov:2015aa} the quasiparticle band gap and the exciton binding energies can be strongly renormalized, effectively changing the optoelectronic properties of the material. Moreover, the structural properties of the SL TMDCs can have an important impact on their electronic structure, for example, \textit{via} strain induced changes of the band structure \cite{Shi2013,JinSub2015} or by enhancing the interaction with the substrate by introducing an interlayer twist angle \cite{Jin2015}. In the case of SL WS$_2$, it is possible to synthesize single crystal flakes reaching dimensions on the order of $\approx$100~$\mu$m using chemical vapor deposition (CVD) growth techniques, involving molecular growth promoters and standard weakly interacting substrates such as SiO$_2$/Si \cite{Ling2014}. Since such flakes are routinely picked up from SiO$_2$ for transfer to other substrates \cite{Elias2013}, one can freely choose the type of interface for the 2D material and thereby investigate its properties in a wide variety of device or substrate environments. 

Given the interplay of the effects on the electronic properties mentioned above and the potential of assembling diverse sample-substrate systems, direct methods that can spatially resolve the electronic structure of the SL TMDCs are needed, in addition to the available luminescence measurements. So far the thickness-dependent electronic structures of exfoliated MoS$_2$ and WSe$_2$ on SiO$_2$ have been investigated with spatially-resolved angle-resolved photoemission (ARPES) \cite{Jin2013,Yeh2015}. In these studies, both the low energy electron diffraction spots and angle-dependent photoemission features were significantly broadened for SL compared to thicker films. This is caused by the mechanical compliance of the SL films, which conform to the known rough surface of SiO$_2$ \cite{lui2009}, spoiling the momentum resolution of electron scattering and emission probes. Low energy electronic parameters such as effective mass and spin-orbit splitting are therefore not reliably quantifiable from samples on such substrates. Therefore, a secondary motivation of our work is to find sufficiently flat substrates so that the ARPES technique can be applied without loss of momentum resolution, in order to access the true low-energy excitations.

It has been theoretically predicted that by encapsulating SL TMDCs in high-$\kappa$ dielectric materials such as HfO$_2$ in a transistor configuration, the enhanced dielectric screening can lead to renormalization of the electron-hole interaction and the band gap in the materials \cite{latini2015,ryou2016}, in addition to enhanced screening of charged Coulomb impurities and thereby a reduction of the overall carrier scattering rate \cite{locquet2006,Cui2015}. In order to directly study the effect of high-$\kappa$ support materials on the electronic structure of SL WS$_2$ with photoemission, we utilize the transition metal oxides TiO$_2$ and SrTiO$_3$ as model systems. We focus our study on SL WS$_2$ transferred on rutile TiO$_2$(100), which is one of the more simple and stable transition metal oxide surfaces. It has a wide band gap of 3.0 eV and a dielectric constant given by $\kappa = 113$ \cite{Diebold2003,Pang2013}. We compare to the case of SrTiO$_3$(100) where $\kappa = 310$ at room temperature and where both TiO$_2$ and SrO surface terminations are possible \cite{bachelet2009}. We endeavor to determine the electronic structure of these systems using the spatially resolved photoemission techniques x-ray photoemission electron microscopy (XPEEM) and ARPES with the synchrotron beam focused on the order of 30-50~$\mu$m. The TiO$_2$ and SrTiO$_3$ substrates have a smooth surface, which is needed in ARPES experiments where a well-defined surface normal vector is a basic requirement to preserve the angular resolution. Finally, by combining 2D and bulk semiconductors it is possible to increase the efficiency of photocatalytic devices \cite{Sang2014,bernardiextraordinary2013} and to control the separation of charge carriers \textit{via} engineered band offsets \cite{hongultrafast2014,jariwala2016}.

\section{Results and Discussion}

\begin{figure*} [t!]
\begin{center}
\includegraphics[width=1\textwidth]{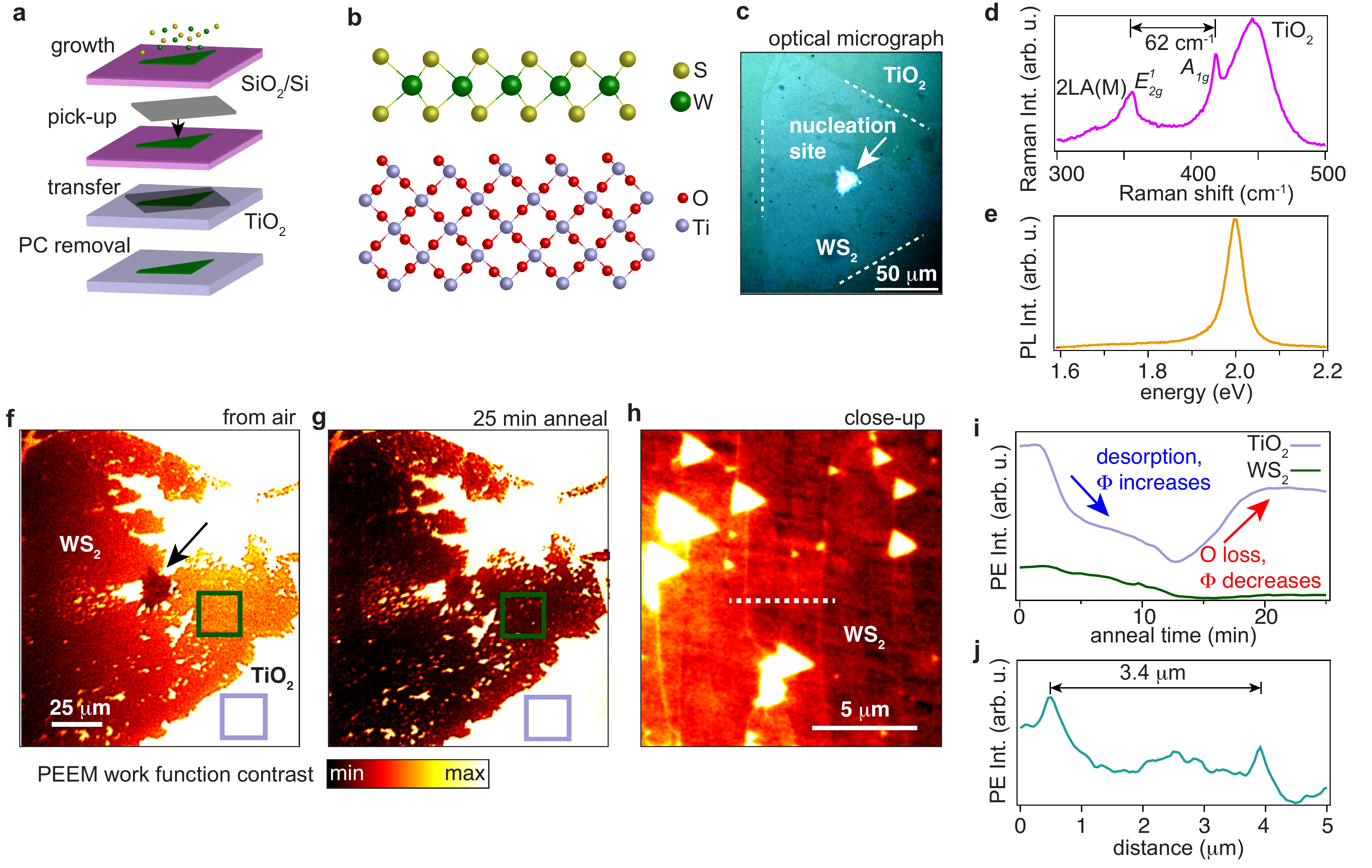}
\caption{Transfer of WS$_2$ on rutile TiO$_2$(100). (a) Scheme of the growth of WS$_2$ on 275 nm SiO$_2$/Si and transfer onto a TiO$_2$(100) surface using a polycarbonate (PC) film. (b) Side-view of a model of the sample-substrate system. (c) Optical microscope image of the transferred SL  WS$_2$ crystal. Dashed lines outline the edges of the crystal. The contrast has been strongly enhanced. (d) Raman spectrum of the crystal in (c) obtained with a 633~nm laser. Characteristic modes are labeled. (e) Photoluminescence spectrum of WS$_2$ on TiO$_2$ obtained with a 488~nm laser excitation. (f)-(h) PEEM images of the area in (c), obtained with a Hg arc discharge lamp for photoexcitation. The images were acquired (f) before and (g)-(h) after 25 minutes of annealing to 600~K. The close-up of the WS$_2$ crystal in (h) was obtained within the area marked by a green square in (f)-(g). (i) Photoemission intensity integrated within the boxed regions in (f)-(g) as a function of time during annealing. (j) Line profile of the photoemission intensity taken across the white dashed line shown in (h). The nucleation site of the WS$_2$ is marked with an arrow in (c) and (f).}
\label{fig:1}
\end{center}
\end{figure*}

A sketch of the entire sample processing route from growth of SL WS$_2$ on 275 nm SiO$_2$/Si using CVD to transfer of SL WS$_2$ on rutile 0.5wt\% Nb doped TiO$_2$(100) is presented in Fig. \ref{fig:1}(a). Further details on the growth and transfer are given in the Methods section and in Refs. \citenum{McCreary2016,Zomer2014}. The simplified model in Fig. \ref{fig:1}(b) summarizes the crystal structures present in the WS$_2$/TiO$_2$ sample. The 2H-stacking type of WS$_2$ with tungsten atoms sandwiched between sulfur layers is supported on the "sawtooth" shaped rutile TiO$_2$(100) surface, such that the bottom sulfur layer in WS$_2$ faces rows of oxygen atoms in TiO$_2$. While the (100) and (110) surfaces of rutile TiO$_2$ have been considered to be the most stable \cite{Muscat1999}, a (1$\times$3) surface reconstruction with a complex structure is known to occur once oxygen vacancies are induced \cite{Lindan2001}. These vacancies appear at the top of the sawtooth profile in close proximity to the WS$_2$ layer. 

We used an optical microscope to locate and inspect the transferred WS$_2$ on TiO$_2$, as shown in Fig. \ref{fig:1}(c). One immediately notices a bright cluster in the center of the image, which corresponds to the nucleation center of the crystal and it likely consists of multi-layered WS$_2$. The silhouette of the triangular shaped SL WS$_2$ crystal around the nucleation site is barely visible as it only gives rise to a contrast enhancement of $\approx1.5$~\% with respect to the bare TiO$_2$ surface. The dashed lines in Fig. \ref{fig:1}(c) have been added as a guide to the eye to locate the edges of the crystal. We observe that the majority of the single triangular WS$_2$ crystal is intact on TiO$_2$ and that it has lateral dimensions on the order of 250~$\mu$m. The Raman spectrum of the transferred WS$_2$ crystal in Fig. \ref{fig:1}(d) reveals the characteristic modes of WS$_2$. These include the in-plane and out-of-plane $E^{1}_{2g}$ and $A_{1g}$ modes as well as the second order longitudinal acoustic mode 2LA(M) that gives rise to a broad shoulder on the $E^{1}_{2g}$ peak. We measure a separation of 62~cm$^{-1}$ between the $E^{1}_{2g}$ and $A_{1g}$ peaks, which is consistent with Raman data from SL WS$_2$ on other substrates \cite{Berkdemir2013,McCreary2016}. The broad peak centered around 450~cm$^{-1}$ is attributed to the $E_g$ mode in rutile TiO$_2$ \cite{Li2015}. The photoluminescence measurement of WS$_2$ on TiO$_2$ in Fig. \ref{fig:1}(e) reveals a single sharp peak around 2.0~eV, which is characteristic of the neutral exciton in transferred WS$_2$ and is typically interpreted as a strong indication of the direct band gap transition at $\bar{K}$ in SL WS$_2$\cite{peimyoo2014}.

The SL WS$_2$/TiO$_2$ crystal is inserted in an ultra-high vacuum (UHV) chamber for PEEM measurements in order to investigate the spatially resolved electronic properties of the material system. The PEEM images shown in Figs. \ref{fig:1}(f)-(h) are obtained using a Hg arc discharge lamp, which has an excitation line centered at $\approx$4.5~eV. Since this energy is close to the work function $\Phi$ of most materials, the main signal derives from emission of secondary electrons around the work function threshold. The photoemission intensity in Figs. \ref{fig:1}(f)-(h) therefore reflects the work function values in the two materials, which have been estimated to be 4.13~eV for the TiO$_2$ substrate \cite{Imanishi2007} and 4.60~eV for SL WS$_2$ \cite{britnellstrong2013}. The lower work function of TiO$_2$ gives rise to a higher intensity and therefore a remarkable contrast difference between the two materials. The PEEM images in Figs. \ref{fig:1}(f)-(g) were obtained before and after 25~min annealing to 600~K, respectively. The image in Fig. \ref{fig:1}(f) presents the WS$_2$/TiO$_2$ directly inserted in the UHV chamber from air after the transfer process outlined in Fig. \ref{fig:1}(a). The anneal time dependence of the photoemission intensity on the WS$_2$ and bare TiO$_2$ areas is tracked in Fig. \ref{fig:1}(i). The initial decrease in intensity in both WS$_2$ and TiO$_2$ is caused by desorption of adsorbates, which increases $\Phi$. The complex shape of these curves and the rise of intensity on the TiO$_2$ part after $\approx15$~min are  associated with annealing induced oxygen vacancy formation in the substrate, which decreases $\Phi$ \cite{Onda2004}. The behavior is seen to saturate after 20~min. 

Fig. \ref{fig:1}(h) presents a closer view of the annealed WS$_2$ crystal. We interpret the narrow vertical lines with a spacing on the order of 3.4~$\mu$m (see line profile in Fig. \ref{fig:1}(j)) as TiO$_2$ intensity that is escaping through cracks in the WS$_2$ film. These develop due to strain relief during annealing and cooling cycles in the transfer process. Moreover, slight intensity variations are visible between cracks, which could arise from very small changes in the coupling to the substrate and thereby work function, as observed for epitaxial graphene in low energy electron microscopy (LEEM) experiments \cite{Diaye2009}. Triangular pits in the WS$_2$ crystal appear to be pinned at the cracks, as seen in Fig. \ref{fig:1}(h). These are consistent with oxidatively etched triangular pits observed in MoS$_2$, which have been speculated to occur from surface diffusion of chemisorbed oxygen that reacts with defect sites \cite{Zhou2013}.

\begin{figure*} [t!]
\begin{center}
\includegraphics[width=0.7\textwidth]{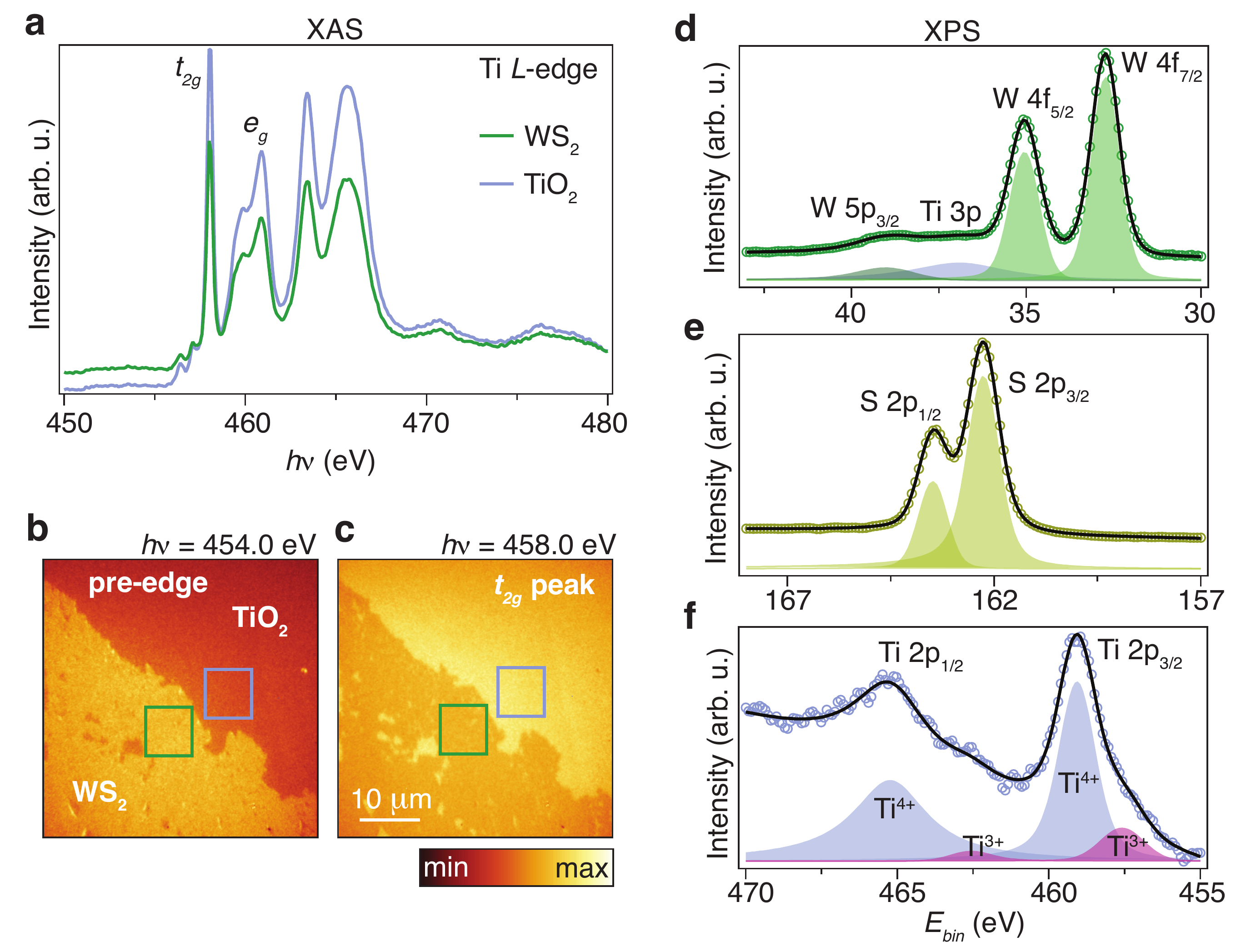}
\caption{Chemical state of the  WS$_2$/TiO$_2$ interface probed by x-ray absorption (XAS) and core level spectroscopy (XPS) using XPEEM. (a) XAS spectrum from the Ti $L$-edge obtained by integrating the secondary electron derived photoemission intensity within the boxed regions in (b)-(c). (b)-(c) XPEEM images in the (b) pre-edge region and (c) at the $t_{2g}$ peak energy. (d)-(f) XPS spectra from the (d) W 4f, (e) S 2p and (f) Ti 2p binding energy regions. Markers are data points and lines are fit results of fits to Doniach-Sunjic line shapes. The shaded peaks correspond to each fitted component. The spectra in (d)-(e) were obtained from an interior part of the WS$_2$ crystal and the spectrum in (f) was measured on the bare TiO$_2$.}
\label{fig:2}
\end{center}
\end{figure*}

Spatially resolved x-ray absorption spectroscopy (XAS) measurements of the Ti L-edge are presented in Figs. \ref{fig:2}(a)-(c). The XAS spectra are sensitive to the chemical composition of the TiO$_6$ octahedral network in the substrate, thereby enabling us to investigate the influence of the WS$_2$ flake on these. The data were acquired with XPEEM by recording the secondary electron signal as a function of photon energy from the sample in a region where the WS$_2$ crystal edge is visible. The Ti L-edge spectrum shown in Fig. \ref{fig:2}(a) was obtained by integrating the photoemission intensity in the boxed regions shown in the XPEEM images in Figs. \ref{fig:2}(b)-(c) for each measured photon energy. We are thus able to extract the XAS spectrum from a specific area on the sample. The Ti L-edge shows the characteristic doublet of peaks arising from the crystal field splitting in TiO$_2$. Within each doublet we identify the so-called $t_{2g}$ and $e_{g}$ peaks, which correspond to the $\pi$- and $\sigma$-states formed by the Ti 3d and O 2p orbitals, respectively \cite{Groot1990}. We find an overall reduction of intensity within the Ti L-edge on the WS$_2$ part due to absorption of secondary electrons in the WS$_2$ crystal. This gives rise to a significant contrast difference between TiO$_2$ and WS$_2$ as seen in the comparison between the images obtained in the pre-edge region and at the $t_{2g}$ peak energy in Figs. \ref{fig:2}(b)-(c). The ratio between the $e_g$ and $t_{2g}$ peaks is nearly identical on the bare TiO$_2$ substrate and on the WS$_2$ part. This implies that the $\sigma$-interaction is uniform across bare TiO$_2$ and under the WS$_2$ \cite{Thomas2007}, such that the annealing- and beam-induced- oxygen vacancy density is similar in the two parts of the sample and there is no apparent chemical interaction between TiO$_2$ and WS$_2$.

\begin{figure*} [t!]
\begin{center}
\includegraphics[width=0.9\textwidth]{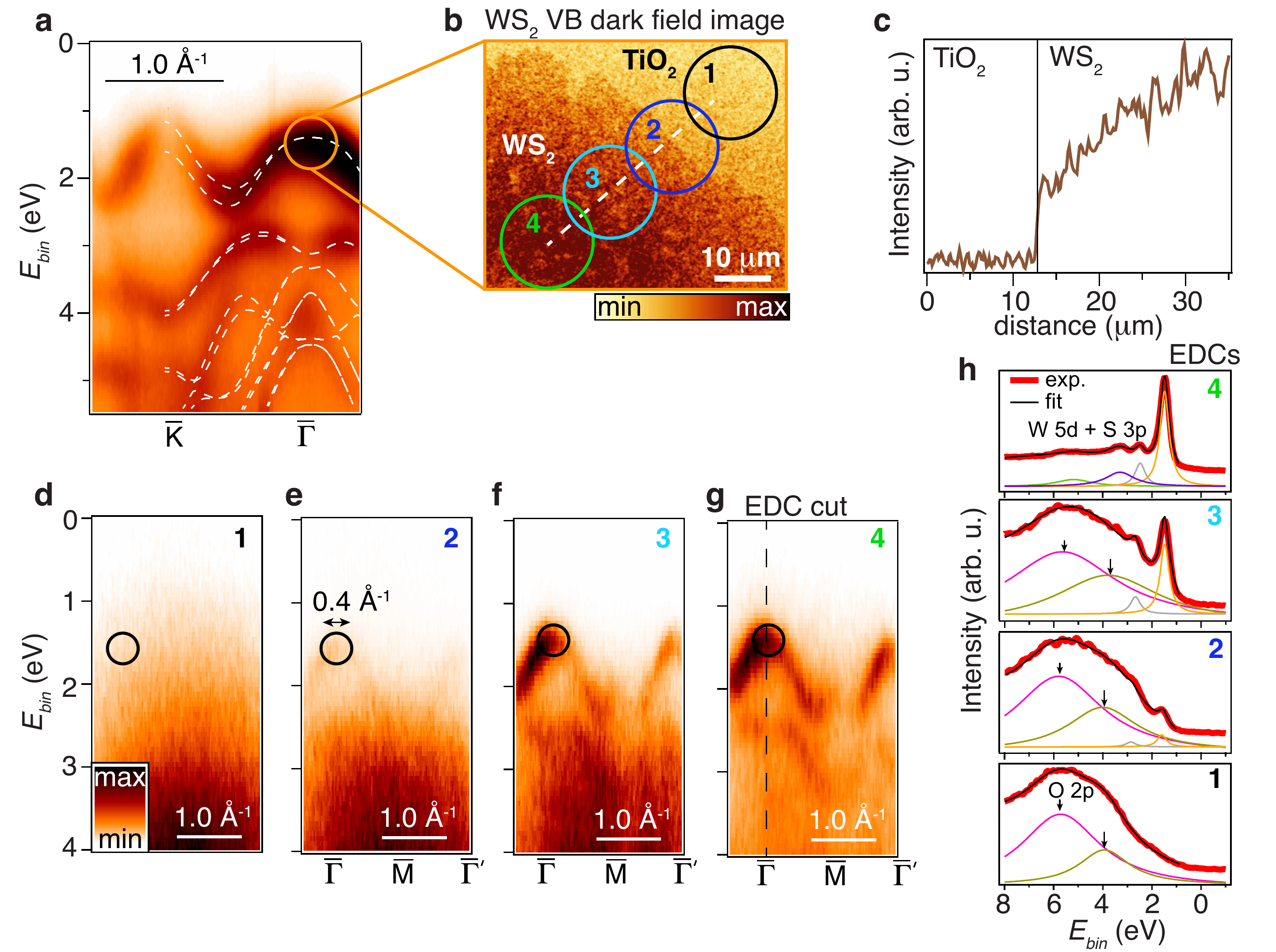}
\caption{Spatial variation  of the WS$_2$ and TiO$_2$ electronic band structures measured with XPEEM. (a) Detailed ARPES spectrum in the $\bar{\Gamma}-\bar{K}$ direction of the WS$_2$ BZ from the circular region labeled 4 in (b). The overlaid white dashed lines represent the calculated band structure for freestanding SL WS$_2$, which was obtained from Ref. \citenum{zhugiant2011}. (b) Dark field image on the upper VB of WS$_2$ around $\bar{\Gamma}$. (c) WS$_2$  VB intensity across the dashed line in (b). The intensity has been normalized to the signal from the TiO$_2$ substrate. (d)-(g) Spatially-resolved ARPES measured within the correspondingly numbered circular regions in (b). The spectra were collected along the WS$_2$ $\bar{\Gamma}-\bar{M}-\bar{\Gamma}^{\prime}$ high symmetry direction. (h) EDC cuts binned $\pm0.1$~\AA$^{-1}$ around $\bar{\Gamma}$ for the data in (d)-(g). The vertical dashed line in (g) illustrates the EDC cut direction. Fit results to multiple Lorentzian line shapes on a parabolic background are displayed for each EDC along with the individual Lorentzian components. The arrows mark the peaks fitted to the O 2p band. The circle around $\bar{\Gamma}$ in (a) and (d)-(g) demarcates the contrast aperture used to acquire the dark field image in (b). The numerical size of the aperture is provided in (e), as shown \textit{via} the double-headed arrow.}
\label{fig:3}
\end{center}
\end{figure*}

Area selective x-ray photoelectron spectroscopy (XPS) data for the W 4f, S 2p and Ti 2p core levels are presented in Figs. \ref{fig:2}(d)-(f). The photoelectrons were measured in a field of view (FOV) of 15~$\mu$m on the sample using an aperture in the center of the energy dispersive prism of the microscope. The spectra were acquired using an energy filtering entrance slit and by projecting the back focal plane onto the detector. See Methods section and Refs. \citenum{Fujikawa2009,Tromp2009} for further details of XPS and ARPES acquisition modes with XPEEM. The W 4f and S 2p binding energy regions in Figs. \ref{fig:2}(e)-(f) were measured in the interior part of the WS$_2$ crystal. For both the W 4f and S 2p regions we observe the expected peaks and no additional components due to \textit{e.g.} tungsten-oxides remaining from the CVD synthesis  \cite{McCreary2016}, or due to an interaction with oxygen in the TiO$_2$ substrate. The Ti 2p spectrum in Fig. \ref{fig:2}(f) was recorded with the aperture on the bare TiO$_2$ substrate. A total of four peaks is used to fit the spectrum to take into account the coexistence of both Ti$^{4+}$ and Ti$^{3+}$ charge states. The latter appears due to oxygen vacancies, which have formed during the initial annealing of the sample and the continuous exposure to the synchrotron beam.

The WS$_2$/TiO$_2$ valence band (VB) region is investigated in Fig. \ref{fig:3}. We measure the electronic band structure using XPEEM with a photon energy of 45~eV and a FOV of 15~$\mu$m on the sample, as described for the XPS measurements above. A spectrum measured in the interior part of the WS$_2$ crystal along the $\bar{\Gamma}-\bar{K}$ symmetry direction of the WS$_2$ BZ is presented in Fig. \ref{fig:3}(a). We identify the characteristic VB of SL WS$_2$ with the global VBM situated at $\bar{K}$ and not at $\bar{\Gamma}$ as in multilayer WS$_2$ \cite{zhugiant2011}. This also implies that our transferred SL WS$_2$ has a direct band gap at $\bar{K}$. The remainder of the dispersing bands all derive from the W 5d and S 3p states that make up the VB of WS$_2$ \cite{zhugiant2011}. We do not observe any changes in the band structure such as hybridization or broadening effects as the WS$_2$ bands cross into the TiO$_2$ VB continuum around a binding energy of 2.5~eV. This is most clearly seen \textit{via} the excellent agreement between the calculated bands of free-standing SL WS$_2$ shown as the white dashed lines in Fig. \ref{fig:3}(a) and the measured band structure.

In order to determine how the VB of WS$_2$ varies spatially we measure an energy- and momentum-filtered image in real space, which corresponds to the dark field imaging mode of XPEEM \cite{Mentes2012}. In the dark field image in Fig. \ref{fig:3}(b) we restrict the measured electrons to originate from the local VBM at $\bar{\Gamma}$ in WS$_2$ using a contrast aperture (see circled region in Figs. \ref{fig:3}(a) and \ref{fig:3}(d)-(g) for a sketch of the contrast aperture) and therefore obtain a higher intensity on the WS$_2$ crystal than on the substrate. However, the line scan of the dark field intensity in Fig. \ref{fig:3}(c) reveals a decrease in VB intensity around the edge of the WS$_2$ crystal.

This behavior is studied in detail \textit{via} the dispersion plots along the $\bar{\Gamma}-\bar{M}-\bar{\Gamma}^{\prime}$ direction of the WS$_2$ BZ in Figs. \ref{fig:3}(d)-(g) and the energy distribution curves (EDCs) binned $\pm0.1$~\AA$^{-1}$ around the $\bar{\Gamma}$-point in Fig. \ref{fig:3}(h). Such EDC cuts are vertical intensity profiles of the photoemission intensity, as shown \textit{via} the dashed line in Fig. \ref{fig:3}(g). These data were acquired with a selective area aperture corresponding to the circled regions in Fig. \ref{fig:3}(b). We observe that the bare TiO$_2$ electronic structure (see Fig. \ref{fig:3}(d)) with oxygen 2p bands peaking around binding energies of 6~eV and 4~eV coexist with WS$_2$ VB features in regions 2 and 3 near the edge of the flake (see EDCs 1-3 in Fig. \ref{fig:3}(h)). This is expected for the edge region probed in Fig. \ref{fig:3}(e) since both bare TiO$_2$ and WS$_2$ is included in the measured FOV. Since the O 2p bands of TiO$_2$ are not completely attenuated by the WS$_2$ in Fig. \ref{fig:3}(f) and EDC 3 in Fig. \ref{fig:3}(h) the WS$_2$ flake must contain nano-sized pinholes, which could have been etched like the larger triangular holes seen in Fig. \ref{fig:1}(g). Note that in the interior of the flake where EDC 4 in Fig. \ref{fig:3}(h) was acquired we exclusively observe peaks that correspond to the W 5d and S 3p states seen in Fig. \ref{fig:3}(a). 

Small, rigid binding energy shifts on the order of 100~meV are observed between the WS$_2$ related peaks in EDC 2-4 in Fig. \ref{fig:3}(h), which implies a spatial doping profile. This is likely caused by remaining polymer residue from the transfer process \cite{Suk2013}. Additionally, changes are known to occur in the structural and chemical composition around the edge of triangular synthetic 2D TMDCs, as the luminescence signal modulates strongly within a range of 0.5-5~$\mu$m from the edge \cite{Bao2015,Gutierrez2013}. The analysis in Figs. \ref{fig:3}(e)-(h) reveals that the WS$_2$ related band structure in the edge region of our transferred flake has similar dispersion and line widths as the band structure measured in the interior part. For example, the full width at half maximum (FWHM) of the Lorentzian line shape fitted to the top-most VB state remains on the order of 550~meV through EDCs 2-4 in  Fig. \ref{fig:3}(h). Thus, the actual band structure does not seem to vary strongly from the edge to the interior of the flake, at least not on the length scale probed here.

\begin{figure*} [t!]
\begin{center}
\includegraphics[width=1\textwidth]{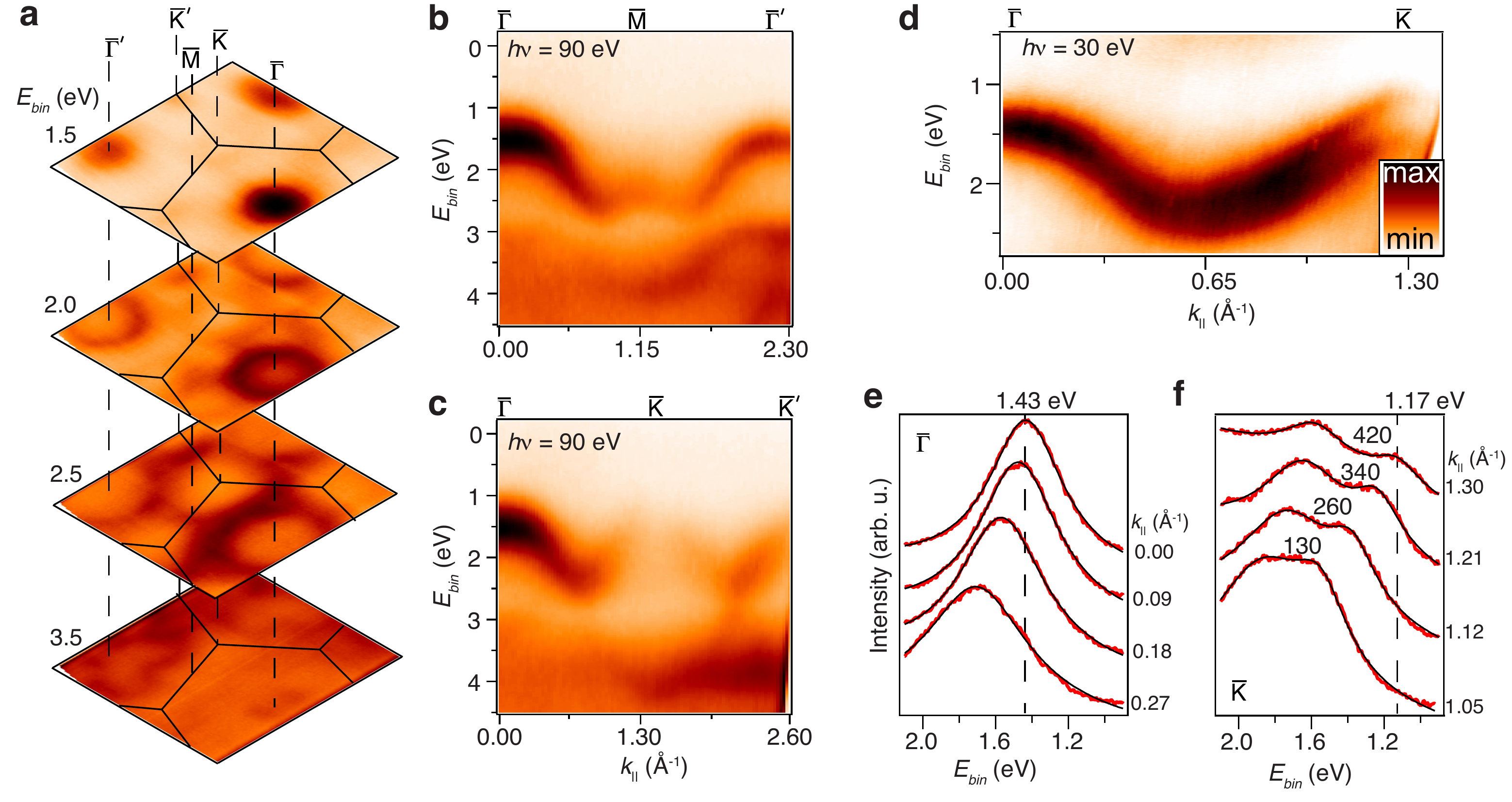}
\caption{ARPES measurements of the VB dispersion and spin-splitting of WS$_2$ on TiO$_2$(100). (a) Constant energy surfaces at the given binding energies. (b)-(c)  Cuts along high symmetry directions of the WS$_2$ BZ. The photon energy in (a)-(c) is 90~eV. (d) Detailed dispersion of the upper VB of WS$_2$ along $\bar{\Gamma}-\bar{K}$ at a photon energy of 30~eV. (e)-(f) EDC analysis of the upper VB around (e) $\bar{\Gamma}$ and (f) $\bar{K}$. In (e) the local VBM of 1.43~eV is shown by a dashed line. In (f) the global VBM of 1.17~eV is shown by a dashed line, and the splitting of the upper VB is given in units of meV for each curve. The standard deviation of the splitting is 10~meV. Red curves are the data and black curves are fits to Lorentzian line shapes on a linear background.}
\label{fig:4}
\end{center}
\end{figure*}

We have performed high resolution ARPES measurements with a hemispherical electron analyzer in order to extract the effective masses and spin-orbit induced splitting of the SL WS$_2$ flake. These measurements have significantly higher energy and angular resolution ($<20$~meV and $<0.1^{\circ}$) but significantly worse spatial resolution ($>30$~$\mu$m) than the measurements discussed above that were acquired using XPEEM.

Well-defined pockets of intensity at $\bar{\Gamma}$ and $\bar{\Gamma}^{\prime}$ are seen in the constant energy surfaces across several BZs of the SL WS$_2$ in the ARPES data in Fig. \ref{fig:4}(a). Faint contours are also discernible around $\bar{K}$ and $\bar{K}^{\prime}$. The observation of distinct WS$_2$ energy contours in multiple BZs ascertain that our flake is a single domain crystal. Cuts of the dispersion along the high symmetry directions spanning two BZs are shown in Figs. \ref{fig:4}(b)-(c). The dispersive features are fully consistent with those observed by XPEEM in Fig. \ref{fig:3}(a) and are attributed to the VB states of SL WS$_2$. The intensity of the bands varies strongly across the BZ and the top of the VB at $\bar{K}$ seems to vanish. This modulation of intensity is a matrix element effect that originates from the change of the W 5d orbital character of the bands from out-of-plane orbitals at $\bar{\Gamma}$ to in-plane orbitals at $\bar{K}$ \cite{Klein2001,miwavander2015}. We find that once we change the photon energy from the 90~eV used in Figs. \ref{fig:4}(a)-(c) to 30~eV we are able to resolve the VBM at $\bar{K}$ along with the spin split states along the $\bar{\Gamma}-\bar{K}$ direction as shown in Fig. \ref{fig:4}(d). Note that the broad line widths observed in the bands in Figs. \ref{fig:4}(b)-(d) are partly attributable to averaging over areas with slightly different doping, which were observed with XPEEM in Fig. \ref{fig:3}(h). 

The detailed measurement of the dispersion in Fig. \ref{fig:4}(d) permits us to analyze EDCs of the VB near $\bar{\Gamma}$ and $\bar{K}$ as shown in Figs. \ref{fig:4}(e)-(f). A single Lorentzian line shape describes the VB around $\bar{\Gamma}$ (see Fig. \ref{fig:4}(e)), which has a maximum at a binding energy of 1.43~eV. The fit leads us to estimate an effective mass given by $m^{\ast}_{\bar{\Gamma}} = (1.55 \pm 0.13)m_e$, where $m_e$ is the free electron mass. At $\bar{K}$ we use two Lorentzian functions to describe the spin-split bands and use the energy difference between the fitted peaks as an estimate of the spin-orbit splitting $\Delta_{SO}$ (see Fig. \ref{fig:4}(f)). We obtain a maximum value given by $\Delta_{SO} = 420$~meV, which diminishes as the band disperses towards $\bar{\Gamma}$. The global VBM is found at a binding energy of 1.17~eV. This places the Fermi level below the middle of the band gap of the material if we assume a quasiparticle gap on the order of 2.9~eV as predicted for SL WS$_2$ \cite{kormanyos2015}. However, since this gap can be strongly renormalized due to screening from the high-$\kappa$ substrate \cite{Ugeda2014,Antonija-Grubisic-Cabo:2015aa}, the actual doping of the WS$_2$ can not be determined from these measurements. We estimate the effective masses of the spin-split bands from the fits and obtain $m^{\ast}_{\bar{K},1} = (0.45 \pm 0.05)m_e$ and  $m^{\ast}_{\bar{K},2} = (0.63 \pm 0.17)m_e$. The value of $\Delta_{SO}$, the observation of the global VBM at $\bar{K}$, and the effective masses are in good agreement with theoretical predictions for free-standing SL WS$_2$ \cite{kormanyos2015}.

\begin{figure*} [t!]
\begin{center}
\includegraphics[width=0.5\textwidth]{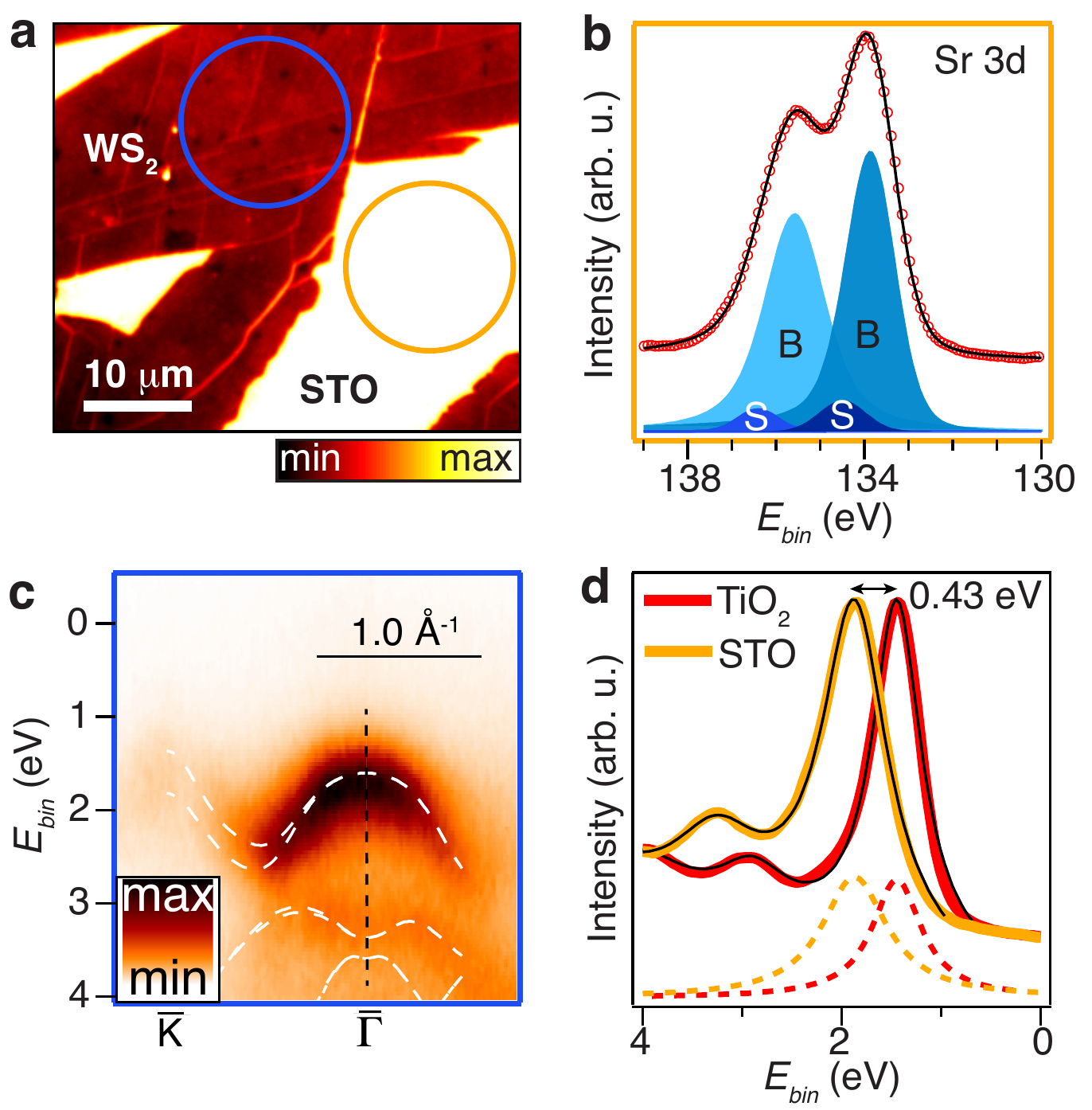}
\caption{PEEM measurements of SL WS$_2$ transferred on SrTiO$_3$ (STO). (a) PEEM image of a transferred flake obtained with a Hg arc discharge lamp. The orange and blue circles demarcate the regions where the measurements in (b) and (c) were performed, respectively. (b) XPS spectrum (markers) from the Sr 3d core level region fitted (lines) with Doniach-Sunjic line shapes. The fitted bulk (B) and surface (S) components are shown \textit{via} the shaded peaks. (c) ARPES spectrum of the SL WS$_2$ VB in the $\bar{\Gamma}-\bar{K}$ direction. White dashed lines are calculated bands of freestanding SL WS$_2$ from Ref. \citenum{zhugiant2011}. (d) EDC at $\bar{\Gamma}$ obtained along the vertical dashed black line in (c). The corresponding EDC for WS$_2$ on TiO$_2$ from Fig. \ref{fig:3}(a) is included for comparison. Fits to Lorentzian peaks are shown \textit{via} the overlaid black curves, and the individual fitted peaks of the top-most VB state are presented \textit{via} the dashed curves. A 0.43~eV shift of the VB between the two systems is shown \textit{via} a double headed arrow.}
\label{fig:5}
\end{center}
\end{figure*}

The sample preparation strategy presented in Fig. \ref{fig:1}(a) is also utilized to transfer SL WS$_2$ on a SrTiO$_3$(100) surface. Fig. \ref{fig:5} presents a series of PEEM measurements, which were carried out to understand this more complex example of a transition metal oxide substrate for SL WS$_2$. The transferred flake is imaged using Hg excitation, which provides a strong work function contrast between WS$_2$ and SrTiO$_3$, as seen in Fig. \ref{fig:5}(a). We observe cracks in the flake, similarly as seen for WS$_2$ on TiO$_2$ in Fig. \ref{fig:1}(h). The chemical composition of the SrTiO$_3$ surface following the WS$_2$ transfer process is investigated using XPEEM measurements of the Sr 3d core level region, as shown in Fig. \ref{fig:5}(b). The spectrum was recorded on the bare SrTiO$_3$ surface using a selective area aperture, which is shown \textit{via} the orange circle in Fig. \ref{fig:5}(a). We observe both surface (S) and bulk (B) components, which are indicative of both SrO and TiO$_2$ terminations \cite{bachelet2009}.  Fig. \ref{fig:5}(c) presents a XPEEM measurement of the WS$_2$ VB in the $\bar{\Gamma}-\bar{K}$ direction, which was obtained using a selective area aperture corresponding to the blue circle in Fig. \ref{fig:5}(a). There is a good agreement between the measured VB and the calculated bands of free-standing SL WS$_2$, similarly as seen in Fig. \ref{fig:3}(a) on TiO$_2$, although the WS$_2$ bands appear to be broader in the SrTiO$_3$ case. We quantify this linewidth broadening \textit{via} EDCs at $\bar{\Gamma}$, as shown in Fig. \ref{fig:5}(d). The FWHM of the fitted lorentzian peak of the top-most VB state is 33~\% larger in the SrTiO$_3$ case. The presence of both TiO$_2$ and SrO terminations causes a more inhomogeneous interface between SL WS$_2$ and SrTiO$_3$, which could give rise to scattering and thereby broadening of the measured linewidth. Furthermore, we observe a substantial rigid shift of the VB of 0.43~eV to higher binding energies in the SrTiO$_3$ case, which is likely related to different pinning of the Fermi level in the two substrates.

\section{Conclusion}
In summary, we have interfaced SL WS$_2$ with the transition metal oxide substrate rutile TiO$_2$(100) and observe \textit{via} XPEEM and ARPES measurements of the electronic structure that the substrate has a very minor influence on the SL WS$_2$ VB, which strongly resembles the VB of a freestanding layer. Previous ARPES studies have focused on epitaxial SL WS$_2$ grown on graphite \cite{Klein2001} or on metal surfaces \cite{Dendzik2015}. The top-most VB of SL WS$_2$ did not exhibit any interaction with the graphite substrate \cite{Klein2001}, but for SL WS$_2$ grown on Au(111) strong hybridization effects were observed between the VB at $\bar{\Gamma}$ and the bulk metal bands \cite{Dendzik2015}. The out-of-plane W 5d and S 3p orbitals that constitute the VB at $\bar{\Gamma}$ are particularly sensitive towards the environment. In our case we do not observe any hybridization effects at $\bar{\Gamma}$ or in the WS$_2$ states that disperse into the VB continuum of the TiO$_2$. Additionally, we do not observe strong distortions of the VB bandwidth or the effective masses at $\bar{K}$, which is in contrast to the related materials MoS$_2$ and WSe$_2$ exfoliated on SiO$_2$ \cite{Jin2013,Yeh2015}. In our case the smooth substrate with a high dielectric constant ensures that distortions in the measured band structure from structural corrugations and charged impurities are minimized. Our results demonstrate that a large spin-orbit splitting, small effective masses and a shift of the VBM to the $\bar{K}$-point, which implies a direct quasiparticle band gap in the material, persist in CVD grown SL WS$_2$ transferred on TiO$_2$. We have shown that  SL WS$_2$ transferred on the more complex oxide SrTiO$_3$ is characterized by broader electronic states, which may be related to SrO and TiO$_2$ segregation at WS$_2$-SrTiO$_3$ interface. Furthermore, a large shift of the SL WS$_2$ VB of 0.43~eV on SrTiO$_3$ compared to TiO$_2$ shows that the choice of substrate strongly influences the band positions in SL TMDCs. Our results imply that the transfer of SL TMDCs on transition metal oxide substrates offers an alternative method for engineering the properties of 2D semiconductors, for example by interfacing the 2D materials with different high-$\kappa$ materials and thereby introducing a means to control the screening and band alignments in such systems. 

\section{Materials and Methods}

\subsection{WS$_2$ synthesis on SiO$_2$}
Synthesis of monolayer WS$_2$ is performed at ambient pressure in a 2-inch diameter quartz tube furnace on Si/SiO$_2$ (275 nm) substrates \cite{McCreary2016}. Prior to use, all Si/SiO$_2$ substrates undergo a standard cleaning procedure consisting of; (i) ultrasonication in acetone, (ii) ultrasonication in isopropanol, (iii) submersion in Piranha etch (3:1 mixture of H$_2$SO$_4$:H$_2$O$_2$) for approximately 2 hours and (iv) thorough rinsing in demineralized water. A quartz boat containing ~1~g of WO$_3$ powder was positioned at the center of the furnace. Two Si/SiO$_2$ (275 nm) wafers are positioned face-down, directly above the oxide precursor. The upstream wafer contains perylene-3,4,9,10-tetracarboxylic acid tetrapotassium salt (PTAS) seeding molecules, while the downstream substrate is untreated. The hexagonal PTAS molecules are carried downstream to the untreated substrate and promote lateral growth of WS$_2$ \cite{Ling2014}. A separate quartz boat containing sulfur powder is placed upstream, outside the furnace-heating zone. Pure argon (100 sccm) is used as the furnace heats to the target temperature. Upon reaching the target temperature of 825 $^{\circ}$C, 10 sccm H$_2$ is added to the Ar flow and maintained throughout the 10 minute soak and subsequent cooling.

\subsection{WS$_2$ transfer to transition metal oxides}
The CVD grown SL WS$_2$ samples on SiO$_2$ are transferred onto 0.5wt\% Nb-doped rutile TiO$_2$(100) and 0.5wt\% Nb-doped SrTiO$_3$(100) substrates purchased from Shinkosha Co., Ltd. The average surface roughness of the substrates is better than 10~\AA ~according to the vendor. TiO$_2$ was cleaned by ultrasonication in acetone and then ethanol, and SrTiO$_3$ was cleaned by buffered hydrofluoric acid etching followed by ultrasonication in acetone and then ethanol.  A polymer-based pick-up technique was used for the transfer process \cite{Zomer2014}. For this, a 5$\times$5$\times$1 mm piece of polydimethylsiloxane (PDMS) stamp is attached on a glass slide. Separately, a polycarbonate (PC) film is prepared on another glass slide by putting a few drops of PC solution and spreading it across the slide (by quickly sliding another glass slide over it). The PC film is peeled off from the glass slide using a clear tape and then attached onto the PDMS stamp. The prepared PC/PDMS glass slide is mounted facing down on a home built transfer tool and carefully lowered to bring in contact with the WS$_2$ on SiO$_2$ substrate, which is held by vacuum on a heating stage. The sample stage is heated to 90 $^{\circ}$C to soften the PC polymer on the WS$_2$ flake and then the polymer is retracted slowly by cooling down the stage.  On retracting, the polymer picks up the WS$_2$ flake from the SiO$_2$ substrate. Then, the PC film carrying the WS$_2$ flake is dropped onto TiO$_2$ or SrTiO$_3$ by melting the polymer at 150~$^{\circ}$C. The PC polymer is removed by rinsing the substrate in chloroform for about 10-15 minutes. During transfer, the whole process is closely monitored through a microscope. After transfer of the WS$_2$ flake on TiO$_2$ or SrTiO$_3$ the sample is annealed in a UHV chamber at 320~$^{\circ}$C for further cleaning.

\subsection{Optical microscopy, Raman and Photoluminescence Spectrocsopy}
A Nikon measuring microscope MM-40 was used for optical imaging of transferred WS$_2$. The Raman spectrum was acquired in ambient conditions with a Renishaw InVia Raman microscope using $E_{laser} = 1.96$~eV. The laser was focused on the WS$_2$ sample with a 50x objective, and a low laser power density ($<$100$\mu$W) was used to avoid heating/creating defects in the sample. Photoluminescence spectroscopy measurements were acquired at room temperature in ambient conditions using a commercial Horiba LabRam confocal spectrometer. A 50x objective was used to focus the 488~nm laser beam to a spot of $~2$~$\mu$m diameter. The laser power at the sample was below 60~$\mu$W.

\subsection{X-ray photoemission electron microscopy (XPEEM)}
The XPEEM measurements were carried out using a SPECS FE-PEEM P90 system installed at the MAESTRO facility at beamline 7.0.2 of the Advanced Light Source (ALS) in Berkeley, CA, USA. An extraction voltage of -5~kV was applied to the sample while an objective lens 1.5~mm away from the sample was kept at ground potential in order to extract photoexcited electrons. Work function imaging was achieved using a Hg arc discharge lamp with an ultraviolet spectrum peaked at $\approx$4.5 eV. XPS core level data were acquired using synchrotron radiation with photon energies of 130~eV, 230~eV, 260~eV and 560~eV for the W 4f, Sr 3d, S 2p and Ti 2p regions, respectively. ARPES data were collected using synchrotron radiation with a photon energy of 45~eV. For ARPES and XPS measurements we used a 15~$\mu$m FOV circular aperture in the electron beam path to select an area of interest on the sample. The energy filtering was achieved using a magnetic prism with a motorized entrance slit with variable slit sizes \cite{Fujikawa2009,Tromp2009}. For ARPES the total energy and momentum  resolution were on the order of 250~meV and 0.03~\AA$^{-1}$, respectively. The binding energy scaling in the ARPES and XPS data was determined by acquiring the data at two different sample voltages offset by 5~V, and the absolute binding energies were determined by measuring the secondary electron cut-off. Microscopy in XAS mode was achieved by filtering the secondary electrons with a contrast aperture in the diffraction plane of the microscope. Dark field imaging on the WS$_2$ VB was done by placing the contrast aperture on the local VBM at $\bar{\Gamma}$. During XPEEM measurements the sample was kept at room temperature.

\subsection{High resolution angle-resolved photoemission spectroscopy (ARPES)}
High resolution ARPES measurements were carried out using the $\mu$ARPES end station equipped with a hemispherical VG Scienta R4000 analyzer at the MAESTRO facility. The lateral size of the synchrotron beam was estimated to be between 30-50~$\mu$m. We used photon energies of 90~eV for overview scans of the BZ and 30~eV for detailed scans of the WS$_2$ VB. The total energy and momentum resolution were better than 20~meV and 0.01~\AA$^{-1}$, respectively. The sample was kept at 85~K during measurements. 

\begin{acknowledgement}
S. U. acknowledges financial support from the Danish Council for Independent Research, Natural Sciences under the Sapere Aude program (Grant No. DFF-4090-00125). R. J. K. is supported by a fellowship within the Postdoc-Program of the German Academic Exchange Service (DAAD). D. S. acknowledges financial support from the Netherlands Organisation for Scientific Research under the Rubicon Program (Grant 680-50-1305). The Advanced Light Source is supported by the Director, Office of Science, Office of Basic Energy Sciences, of the U.S. Department of Energy under Contract No. DE-AC02-05CH11231. This work was supported by IBS-R009-D1. The work at Ohio State was primarily supported by NSF-MRSEC (Grant DMR-1420451). Work at NRL was supported by core programs and the NRL Nanoscience Institute, and by AFOSR under contract number AOARD 14IOA018- 134141.
\end{acknowledgement}

\providecommand{\latin}[1]{#1}
\providecommand*\mcitethebibliography{\thebibliography}
\csname @ifundefined\endcsname{endmcitethebibliography}
  {\let\endmcitethebibliography\endthebibliography}{}

\end{document}